\documentclass[twocolumn,superscriptaddress,prl,10pt]{revtex4-1}
\usepackage{verbatim}
\usepackage{braket}
\usepackage{amsmath,amssymb}
\usepackage{tikz}
\usepackage{graphicx}
\usepackage{color}
\usepackage[colorlinks,bookmarks=false,citecolor=blue,linkcolor=red,urlcolor=blue]{hyperref}
\usepackage{times}

\def\tr{\textrm{Tr}}

\def\doi{http://dx.doi.org/}

\newcommand{\be}{\begin{equation}}
\newcommand{\ee}{\end{equation}}

\usepackage{cancel}

\begin{document}

\title{Prethermalization at Low Temperature: the Scent of Long-Range Order} 

\author{Vincenzo Alba}
\affiliation{International School for Advanced Studies (SISSA),
Via Bonomea 265, 34136, Trieste, Italy, 
INFN, Sezione di Trieste}
\author{Maurizio Fagotti}
\affiliation{Departement de Physique, Ecole Normale Superieure / PSL Research University, 
CNRS, 24 rue Lhomond, 75005 Paris, France}


\begin{abstract} 
Non-equilibrium time evolution in isolated many-body  quantum systems generally results in thermalization. However, the relaxation process can be very slow, and quasi-stationary non-thermal plateaux are  often observed at intermediate times.  The  paradigmatic example is a quantum quench in an integrable model with weak integrability breaking; for a long time,  the state can not escape the constraints imposed by the approximate integrability. We unveil a new mechanism of prethermalization, based on the presence of a symmetry of the pre-quench Hamiltonian, which is spontaneously broken at zero temperature,  
and is explicitly broken by the post-quench Hamiltonian. The typical time scale of the phenomenon is proportional to the thermal correlation length of the initial state, which diverges as the 
temperature is lowered.  
We show that  the prethermal quasi-stationary  state can be approximated by a mixed state that violates cluster decomposition property.  
We consider two examples: the transverse-field Ising chain, where the full time evolution is computed analytically, and the (non integrable) ANNNI model, which is investigated  numerically. 

\end{abstract}


\maketitle

\paragraph*{Introduction.}

\begin{figure*}[t]
\includegraphics*[width=0.93\linewidth]{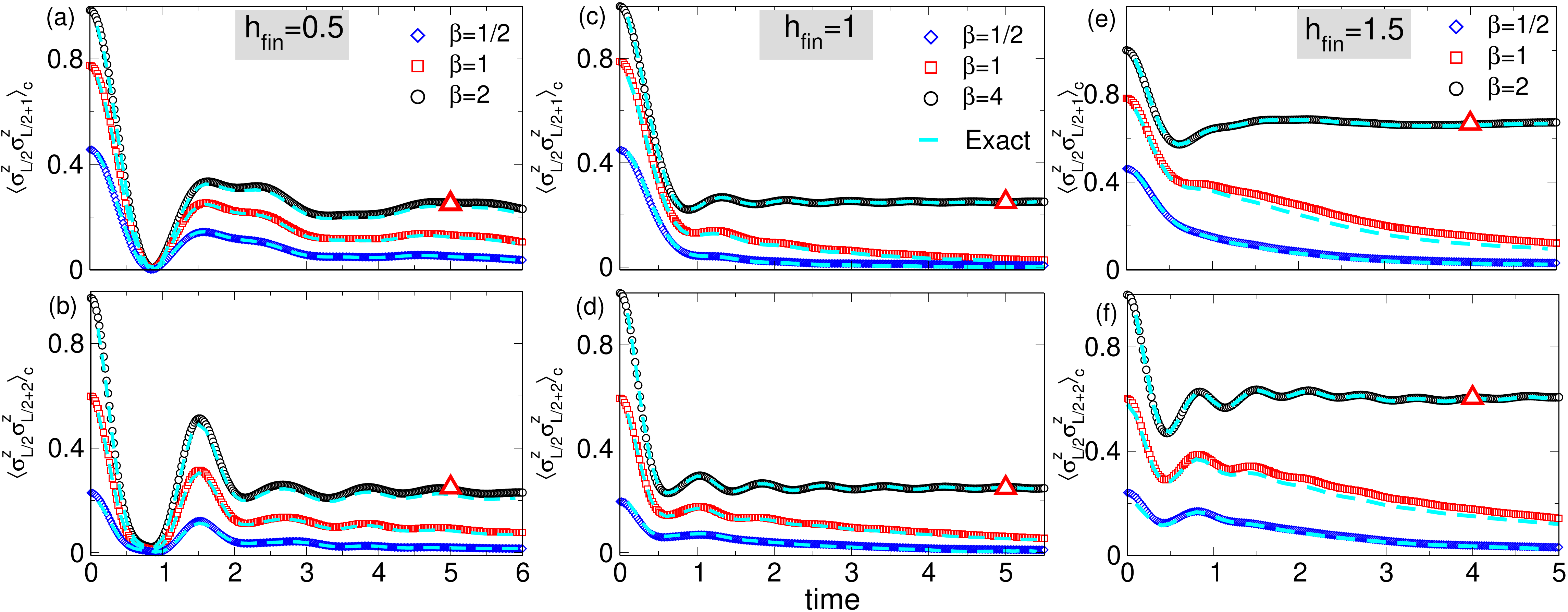}
\caption{ 
Post-quench dynamics 
 of the connected spin-spin correlators in the Ising spin chain. 
  The initial thermal ensemble 
 of the classical Ising model ${H}^{{Is}}_{x}$ at inverse temperature $\beta$ is evolved under the rotated 
 Ising Hamiltonian ${H}^{Is}_{z}$ with transverse field $h_{f}$. The different panels show 
 $\langle \sigma^z_{L/2}\sigma^z_{L/2+r}\rangle_c$, with $r=1,2$ for several 
 quenches, as a function of the time after the quench. The data are 
 tDMRG results for a chain with 
 $L=32$ sites. 
 In all panels the error bars are often smaller than the symbol size. 
 The dashed lines are exact analytical results. 
 The triangles are the expected values for the prethermalization plateaux. 
In the infinite time limit the GGE prediction $\langle \sigma^z_{L/2}
 \sigma_{L/2+r}^z\rangle_c=0$ is recovered. 
}
\label{fig1}
\end{figure*}

The dynamics of isolated many-body quantum systems have attracted a lot of attention~\cite{exp,EF:review,quench,GE15, 
SC:CD}.  
After quenching a global Hamiltonian parameter, the physical properties of these systems become eventually stationary, and can be described using the language of statistical physics. It is now common to say that the system has thermalized when the observables of interest can be described by a Gibbs ensemble (or by a grand canonical ensemble), notwithstanding the associated temperature not being generally equal to the actual temperature of the system.  
Generic translationally invariant models are expected to thermalize~\cite{ETH}, with the noteworthy exception of integrable models, which have infinitely many (quasilocal) conservation laws that constrain the dynamics~\cite{charges}; the Gibbs ensemble must then be replaced by the so-called generalized Gibbs ensemble (GGE)~\cite{rigol-2007}. 

Much less is known of the relaxation process.  This can be extremely slow and, often, a distinct quasi-stationary behavior emerges at intermediate times. This was called \emph{prethermalization}~\cite{preT}, as it was originally observed in models that were supposed to thermalize. The typical example is non-equilibrium time evolution in an integrable model with weak integrability breaking~\cite{prethermalization,essler-2014,babadi-2015}. If the integrability-breaking interactions are weak enough, in a large time window their effects are minor and 
the expectation values of the most local observables approach quasi-stationary plateaux. Remarkably, these can be approximately described by a deformed GGE~\cite{essler-2014}. 
This phenomenon has been observed in several experiments on ultra-cold bosonic gases~\cite{preT:exp}.
More recently, it has been shown that similar phenomena can occur even if the perturbation preserves integrability~\cite{prerelaxation}. 

In this Letter we unveil a novel mechanism of prethermalization in systems where long-range order is present only at zero temperature. 
These systems are characterized by a correlation length which diverges upon lowering the temperature
~\cite{sachdev-book,chakravarty-1989}; however, the physical ground state has a finite correlation length.   
There is a wide class of models with this property, especially in 1D and 2D; there, at finite temperature continuous symmetries are generally unbroken~\cite{Mermin,Hohenberg}, and, in 1D in particular, this is expected to be true also for discrete symmetries.
Relevant examples are  the 1D quantum Ising ferromagnet~\cite{sachdev-book} and the 2D quantum Heisenberg antiferromagnet~\cite{chakravarty-1989}.  We will show that, at low temperature, a quench with a Hamiltonian that breaks the symmetry of the state gives rise to an unusual prethermalization. 
\paragraph{Quench protocol.}
Let $H_0$ be the pre-quench Hamiltonian, and 
$\ket{{\rm GS}_j}$  
the ground states of $H_0$ which can be selected when the symmetry is broken. 
We prepare the state at equilibrium at inverse temperature $\beta$, and quench the Hamiltonian by adding a local perturbation that explicitly  breaks the symmetry. The density matrix at time $t$ reads as
\be\label{eq:rhot}
\rho(t)=Z^{-1} e^{-i H t} e^{-\beta H_0}e^{i H t}\, ,
\ee
where $H$ is the post-quench Hamiltonian and $Z$ is the partition function. We are interested in the behavior of local observables, so we can focus on the reduced density matrix $\rho_A(t)=\tr_{\overline A}[\rho(t)]$ of some subsystem $A$ ($\overline A$ is the complement of $A$).  Generally, at late times, $\rho_A(t)$ becomes time-independent and can be described by a Gibbs ensemble (if not explicitly stated otherwise, we consider a generic, i.e., non-integrable model):
\be
\lim_{t\rightarrow\infty }\rho_A(t)={\tilde Z}^{-1}\tr_{\overline A}\bigl[e^{-\tilde \beta H}\bigr]\, ;
\ee
here $\tilde \beta$ is fixed by the conservation of energy and can be different from the temperature of the initial state. 
\paragraph{Prethermalization plateaux.} 
For the sake of simplicity, we focus on spin chains. 
At the initial time, if $A$'s length $|A|$ is much smaller than the thermal correlation length $\xi_T$, at the leading order,  $\xi_T$ can be assumed to be infinitely large. In the systems we are considering, this corresponds to the limit $\beta \rightarrow +\infty$. Since the symmetry is unbroken at any finite temperature, this results in replacing the initial state by the following mixed state
\begin{equation}
\label{intro-1}
\lim_{\beta \rightarrow +\infty} Z^{-1}e^{-\beta H_0}=\frac{1}{n}\sum\nolimits_{j=1}^n \ket{{\rm GS}_j}
\bra{{\rm GS}_j}\, ,
\end{equation}
where we assumed that the symmetry can be broken in $n$ distinct ways. 
We focus on situations in which $\braket{GS_j|H|GS_j}$ explicitly depends on $j$, which is possible only because $H$ does not share the symmetry of $H_0$. 
Let us then consider the time evolution of an observable $\mathcal O$, acting nontrivially only on $A$. Ref.~\cite{bravyi06} showed that 
$\mathcal O(t)=e^{i H t}\mathcal O e^{-i H t}$ can be approximated by a local operator with range $|A|+2 v t+x$, with $x$ a time-independent parameter, keeping the error exponentially small $\sim e^{-x/\xi}$; here $\xi$ is a constant dependent only on $H$, and $v$ is the 
Lieb-Robinson velocity~\cite{LR72}. Thus, the approximation \eqref{intro-1} makes sense also at finite time, provided that $|A|\ll \xi_T-2 v t$; 
we can approximate the density matrix \eqref{eq:rhot} as the incoherent sum of $n$ time-evolving states 
\be\label{eq:cat}
\rho_A(t)\sim \frac{1}{n}\sum\nolimits_{j=1}^n \mathrm{tr}_{\overline A}\bigl[e^{-i H t}\ket{{\rm GS}_j}
\bra{{\rm GS}_j}e^{i H t}\bigr]\, .
\ee 
In generic models, each of these states is expected to locally relax to a thermal ensemble ${\tilde Z_j}^{-1}e^{-\tilde \beta_j H}$, where $\tilde \beta_j$ is fixed by the conservation of energy. By assumption (see below \eqref{intro-1}),  $\tilde \beta_j$ explicitly depends on $j$. 
Let us call $\tau^{(j)}_A$ the  relaxation time~\cite{f:5} of the subsystem $A$ after the quench from $\ket{{\rm GS}_j}$, and denote the maximal time by $\tau_A^{\rm pth}=\max_j \tau^{(j)}_A$. If $\xi_T$ is  sufficiently large (hence, low temperature), in the time window $\tau_A^{\rm pth}<t\ll \frac{\xi_T-|A|}{2v}$, 
$\rho_A(t)$ will be quasi-stationary
\be\label{eq:catGGE}
\rho_A(t)\rightarrow \rho_A^{\rm pth}\equiv \frac{1}{n}\sum\nolimits_{j=1}^n\mathrm{tr}_{\overline A}\bigl[ {\tilde Z_j}^{-1}e^{-\tilde \beta_j H}\bigr]\, .
\ee
Eq.~\eqref{eq:catGGE} reveals a new form of prethermalization, which, as shown below,  differs from prethermalization close to an integrable point in being described by a stationary state without cluster decomposition (CD) property. 
We remark that $\rho_A^{\rm pth}$ is nonthermal because the energy distribution of~\eqref{intro-1} 
is not narrow in the thermodynamic limit, which is necessary for thermalization to 
occur~(\emph{cf.} Appendix A of \cite{EF:review}). This, in turn, reflects that the states appearing in the 
superposition~\eqref{eq:cat} are {\it macroscopically} different. 
\paragraph{CD.}
The CD property states that correlations between local 
operators factorize at large distances, i.e., 
\begin{equation}
\label{cd}
\big\langle{\mathcal O}_\ell{\mathcal O}_m
\big\rangle_c\equiv\big\langle{\mathcal O}_\ell{\mathcal O}_m
\big\rangle-\big\langle{\mathcal O}_\ell\big\rangle
\big\langle{\mathcal O}_m\big\rangle
\stackrel{|\ell-m|\to\infty}{\overrightarrow{\hspace{40pt}}}
0. 
\end{equation}
Here $\langle\cdot\rangle$ denotes the state expectation value and $\langle\cdot\rangle_c$ the connected correlation. 
CD is a fundamental 
property of generic physical states, such as ground states of local Hamiltonians, 
or thermal states~\cite{localT}, and is deeply related to the spontaneous symmetry breaking mechanism~\cite{weinberg-book}. In quench dynamics, the importance of CD was first pointed out in \cite{SC:CD}. 
We show that $\rho^{\rm pth}$ can violate CD by considering the two-point function of the energy density $h_\ell$ ($H=\sum_\ell h_\ell$). We have
\begin{multline}\label{eq:CDproof}
\braket{h_\ell h_m}^{\rm pth}=\sum\nolimits_{j=1}^n\!\!\!\frac{\braket{h_\ell h_m }_j}{n}\xrightarrow{|\ell-m|\rightarrow\infty}\sum\nolimits_{j=1}^n\!\!\!\frac{\braket{h_\ell }_j^2}{n}=\\
(\braket{h_\ell}^{\rm pth})^2+\frac{1}{2n^2}\sum\nolimits_{j,j'=1}^n(\braket{h_\ell}_j-\braket{h_\ell}_{j'})^2\, ,
\end{multline}
where $\braket{\cdot}_j$  and $\braket{\cdot}^{\rm pth}$ denote the expectation values in the 
thermal state with inverse temperature $\tilde \beta_j$, and in $\rho^{\rm pth}$ (\emph{cf}. 
\eqref{eq:catGGE}), respectively. 
In the first line, we used cluster decomposition property of thermal states. 
Since at least two of the $\tilde\beta_j$ are different, the last term of 
\eqref{eq:CDproof} is nonzero, \emph{i.e.},  the stationary state describing 
the prethermal regime does not have CD property. 
As a consequence, the fluctuations of macroscopic quantities, like $|A|^{-1}\sum_{n\in A}h_n$,  
remain finite even for large $|A|$, and they can be used to probe $CD$ violations. 
This is remnant of the pre-quench state having 
long-range order at zero temperature. 
%
As an explicit example, we consider  the 
transverse field Ising chain (TFIC). The Hamiltonian reads as 
\begin{equation}
\label{ising}
{H}^{{Is}}_{x}\equiv-J\sum\nolimits_{i=1}^{L} (\sigma_i^z\sigma^z_{i+1}+h
 \sigma_i^x), 
\end{equation}
with $\sigma^\alpha$ the Pauli matrices, $L$ the chain length, and $h$ an 
external magnetic field. 
We impose periodic boundary conditions ($\sigma_{L+n}^\alpha\equiv \sigma_{n}^\alpha$) and set $J=1$.
The Hamiltonian ${H}_x^{Is}$ is invariant 
under the spin-flip symmetry $\sigma_i^{z,y}\to -\sigma_i^{z,y}$. 
At zero temperature, for $h>1$, the ground state is paramagnetic, while, for $h<1$, is ferromagnetic. 
The two phases are separated by a second order quantum phase transition at $h=1$. Specifically, in both 
regions of the phase diagram and for any {\it finite} size and $h$, the ground state 
$|\psi_0\rangle_{NS}$ 
is in the so-called Neveu-Schwarz sector. This is the sector consisting of the states with $\prod_{i=1}^L\sigma_i^x=1$. 
In the ferromagnetic phase, upon increasing 
$L$, $|\psi_0\rangle_{NS}$ becomes degenerate with the lowest energy state $|\psi_0
\rangle_R$ in the Ramond sector, which consists of the states with $\prod_{i=1}^L\sigma_i^x=-1$.
In the thermodynamic limit spin-flip symmetry is spontaneously broken and one of the two combinations $|\pm\rangle\equiv(|\psi_0
\rangle_{NS}\pm|\psi_0\rangle_R)/\sqrt{2}$ is chosen.  
Crucially, while both $|\psi_0\rangle_{NS}$ and $|\psi_0\rangle_{R}$ 
violate CD,  $|\pm\rangle$ does not; this reflects the stability of the selected 
symmetry-broken state under local perturbations~\cite{weinberg-book}. In particular, for $h=0$ we have $|\psi_0\rangle_{NS}\to(\left|\Uparrow\right\rangle+
\left|\Downarrow\right\rangle)/\sqrt{2}$, and $|\psi_0\rangle_{R}\to(\left|\Uparrow\right
\rangle-\left|\Downarrow\right\rangle)/\sqrt{2}$, with $\left|\Uparrow\right\rangle\equiv
\left|\uparrow\dots\uparrow\right\rangle$ and $\left|\Downarrow\right\rangle\equiv
\left|\downarrow\dots\downarrow\right\rangle$ ($\sigma^z\ket{\uparrow}=\ket{\uparrow}$ and $\sigma^z\ket{\downarrow}=-\ket{\downarrow}$); CD is violated since ${}_{NS(R)}\langle\psi_0|\sigma_i^z
\sigma_j^z|\psi_0\rangle_{NS(R)}=1$, despite ${}_{NS(R)}
\langle\psi_0|\sigma^z_i|\psi_0\rangle_{NS(R)}$ being zero $\,\forall i,j$. On the other hand, at finite temperature the connected two-point function is given by $\langle \sigma^z_1\sigma^z_{1+r}\rangle_c=\tanh^r(\beta)$,  which gives the finite correlation length $\xi_T=-1/\log(\tanh\beta)$.

\paragraph*{Quench details \& method.}

We consider the TFIC~\eqref{ising} in zero field ($h=0$) and prepare the system in a thermal state at inverse temperature $\beta$; this is
described by the density matrix 
\begin{equation}
\label{gibbs}
\rho^{Gibbs}\equiv Z^{-1}e^{-\beta{H}_{x}^{Is}}. 
\end{equation}
The state is let to evolve under the 
ANNNI Hamiltonian 
\begin{equation}
\label{anni-ham}
{H}^{\textsc{an}}_{z}=-\sum\nolimits_{i=1}^{L}(J_1\sigma_{i}^x\sigma_{i+1}^x
+J_2\sigma_i^x\sigma_{i+2}^x+h_{f}\sigma_i^z). 
\end{equation}
Hereafter we set $J_1=1$. 
For $J_2=0$, 
this is the Hamiltonian ${H}^{{Is}}_{z}$ 
of the transverse field Ising chain (note that the direction of the interaction is tilted by $\pi/2$ with respect to \eqref{ising}). 
In that case, by working in the Heisenberg picture, the time evolution of sufficiently simple observables can be obtained analytically.
We focus on the two-point function of the transverse field $\braket{\sigma_i^z\sigma_{i+n}^z}$. The operator $\sigma_i^z\sigma_{i+n}^z$ is mapped to a four fermion operator by the Jordan Wigner transformation $\sigma_\ell^{x,y}=\prod_{j<\ell}(i a_j^y a_j^x) a_\ell^{x,y}$, with $a_j^\alpha$ Majorana fermions satisfying $\{a_j^\alpha,a_\ell^\beta\}=2\delta_{\alpha\beta}\delta_{j \ell}$. For $J_2=0$, time evolution preserves the number of fermions, so the time-evolving operator still consists of four fermions, and can be worked out exactly. The expectation value in the initial state is obtained by extracting the diagonal part of the operator in the $\sigma^z$ basis, which diagonalizes \eqref{gibbs} (for $h=0$).

For generic $J_2$ the model is not integrable, and the dynamics are investigated numerically. 
The procedure is as follows. First, 
we simulate the initial thermal ensemble~\eqref{gibbs}. 
Since the eigenstates of
${H}^{Is}_{x}$ are product states of the form $|\varphi\rangle=\prod_{i=
1}^{{}_L}|s_i\rangle$, with $s_i=\uparrow\vee \downarrow$ , the initial ensemble can be 
conveniently generated using a Metropolis scheme; starting from an arbitrary initial 
state $|\varphi_0\rangle$, a new eigenstate is generated by flipping the spins at 
each site $i$ with probability $\textrm{min}(1,\exp(-\beta(E'-E))$, $E$ being the 
energy of $|\varphi\rangle$, and $E'$ that of the eigenstate with the spin flipped; the 
iteration of this procedure generates the initial thermal ensemble. The ensemble 
average corresponds to the arithmetic mean of the expectation values over the 
sampled eigenstates. Using tDMRG algorithms~\cite{tDMRG}, the dynamics of each state  in the initial ensemble are simulated one by one, and  finally averaged~\cite{f:2}. 

\paragraph*{Results.}
First, we consider the integrable case $J_2=0$.
In the thermodynamic limit, we find that the one-point function of the transverse field is zero at any time, whereas the two-point function can be written as follows
\begin{multline}
\label{corr-an}
\braket{\sigma_1^z\sigma^z_{1+r}}=
\iiint_{-\pi}^\pi\frac{\mathrm d k\mathrm d p\mathrm d q}{(2\pi)^3}
\cos(r q)\\
\times\Bigl[K(q)-K(p+k)\Bigr]
\det\begin{pmatrix}
M_{11}^{p,p-q}&-M_{12}^{p,p-q}\\
M_{21}^{k,k+q}&M_{22}^{k,k+q}
\end{pmatrix}\, ,
\end{multline}
where $K(q)\equiv\frac{\sinh(\gamma)}{\cosh(\gamma)-\cos(q)}$,  $\gamma\equiv2\textrm{
arctanh}(e^{-2\beta})$, and 
\be
\begin{aligned}
M_{11}^{p,q}&=\sin(\varepsilon_p t)\cos(\varepsilon_{q} t)e^{-i\theta_p}-\cos(\varepsilon_p t)\sin(\varepsilon_{q} t)e^{i\theta_{q} }\\
M_{22}^{p,q}&=\sin(\varepsilon_p t)\cos(\varepsilon_{q} t)e^{i\theta_p}-\cos(\varepsilon_p t)\sin(\varepsilon_{q} t)e^{-i\theta_{q} }\\
M_{12}^{p,q}&=\cos(\varepsilon_p t)\cos(\varepsilon_{q} t)+\sin(\varepsilon_p t)\sin(\varepsilon_{q} t) e^{-i\theta_p-i\theta_{q}}\\
M_{21}^{p,q}&=\cos(\varepsilon_p t)\cos(\varepsilon_{q} t)+ \sin(\varepsilon_{q} t)\sin(\varepsilon_p t) e^{i\theta_{q}+i\theta_p}\, .
\end{aligned}
\ee
Here $e^{i\theta_k}=2(e^{i k}-h)/\varepsilon_k$ and $\varepsilon_k=2\sqrt{1+h^2-2 h\cos k}$.
Fig.~\ref{fig1} shows the time evolution of the connected mid-chain correlators $\langle \sigma^z_{L/2}\sigma^z_{L/2+r}\rangle_c$, 
for $r=1$ and $r=2$, after several quenches. 
The  agreement between finite-size tDMRG data (symbols) and  the exact analytic predictions (dashed lines) is very good; this shows that finite-size 
effects are negligible until the largest time simulated. 
The late-time dynamics are described by the GGE with maximal entropy at fixed local integrals of motion. Specifically, for any finite $\beta$, every traceless operator turns out to have zero expectation value. 
Nevertheless, upon increasing $\beta$ and for $r\ll \xi_T$, the correlators exhibit a prerelaxation plateau~\cite{f:4} with $\langle\sigma^z_{L/2}\sigma^z_{L/2+r}\rangle_c\ne0$. This is not described by a standard GGE (Wick's theorem does not apply) but, analogously to \eqref{eq:catGGE},  can be written as the sum of two of them. 
\begin{figure}[t]
\includegraphics*[width=0.93\linewidth]{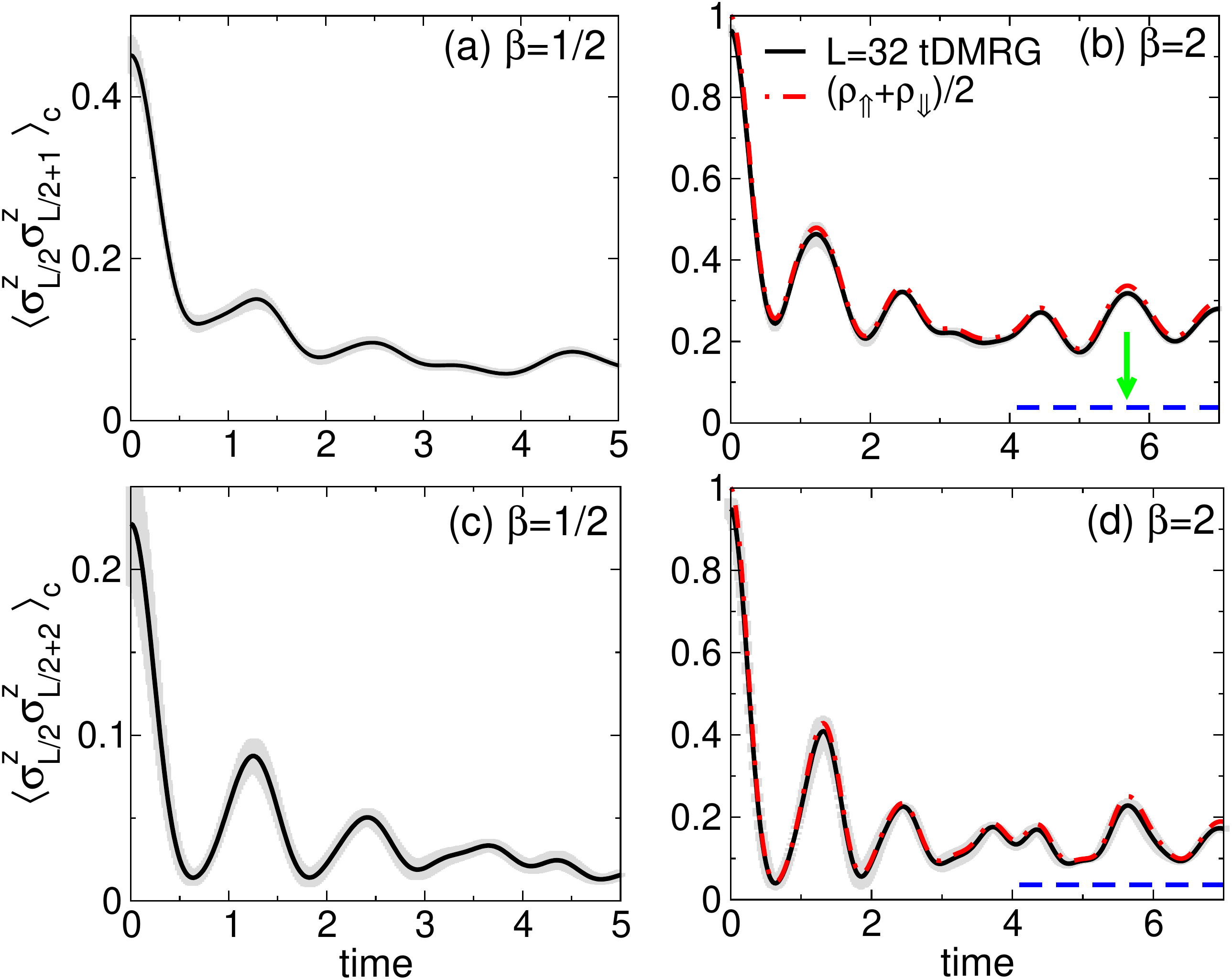}
\caption{
The thermal ensemble of the classical Ising model 
 ${H}^{Is}_{x}$ at inverse temperature $\beta$ is evolved under 
  the ANNNI Hamiltonian ${ H}^{\textsc{AN}}_{z}$ with $J_2=1/2$ and $h_f=1/2$. We show the post-quench dynamics of the connected 
 correlators $\langle\sigma^z_{L/2}\sigma^z_{L/2+r}\rangle_c$ for $r=1$ ((a)(b)) 
 and $r=2$ ((c)(d)). The full lines  are obtained using tDMRG. 
 The shading denotes the statistical error bars. Data are for a chain with $L=32$ sites.  
 In (b)(d) the dash-dotted line is the average of the time evolutions ensuing 
 from $\left|\Uparrow\right\rangle$ and $\left|\Downarrow\right
 \rangle$. The horizontal lines are the expected prethermalization 
 plateaux. 
}
\label{fig2}
\end{figure}
%
The value of the prerelaxation plateau is obtained by taking the limit $\beta\rightarrow\infty$ in \eqref{corr-an}, which gives
\be
\begin{aligned}
\label{pla-an}
\braket{\sigma_1^z\sigma^z_{1+r}}_{c}=&\iint_{-\pi}^\pi
\frac{\mathrm d k\mathrm d q}{(2\pi)^2}
\cos\theta_{k}\cos\theta_qf_r(k,q) \\
f_r(k,q)=&(1+\sin\theta_{k}\sin\theta_q)\sin(r k)\sin(r q)\\
&+\cos\theta_{k}\cos\theta_q(1-\cos(r k)\cos(r q))\, .
\end{aligned}
\ee
The predictions from~\eqref{pla-an} are shown as triangles in Fig.~\ref{fig1}, and are in a very good agreement with the approximate plateaux, visible also in the tDMRG data already for $\beta=2,4$. 

We also identify a prerelaxation scaling limit, where the dynamics of local observables occur in the typical timescale  $(\gamma v)^{-1}\sim \xi_T/v$, where $v=\max_k|\varepsilon'_k|$. 
Specifically, in the limit $vt\gg 1,\gamma r\ll 1$, we find
\be
\label{pla-an}
\braket{\sigma_1^z\sigma^z_{1+r}}_{c}\rightarrow\iint_{-\pi}^\pi
\frac{\mathrm d k\mathrm d q}{(2\pi)^2}
\cos\theta_{k}\cos\theta_qf_r(k,q)e^{-|\varepsilon'_{k}+\varepsilon'_{q}| \gamma t}. 
\ee
For sufficiently large $\gamma v t$, this behaves as $\sim (\gamma v t)^{-1}$, so the relaxation to the GGE turns out to be very slow.  We mention that, if $\gamma r$ is not small, there is an additional contribution, coming from \eqref{corr-an}, that varies on the typical distance $\gamma r$ and time $\gamma v t$~\cite{f:1}; this describes a light-cone dynamics, which is always expected in the space-time scaling limit  $r,v t\rightarrow\infty$, at fixed ratio $r/(vt)$~\cite{EF:review}.   

We now turn to discuss the non-integrable case, considering the post-quench dynamics driven by 
the full ANNNI Hamiltonian~\eqref{anni-ham}. We restrict ourselves to the case $J_2=1/2$ and $h_f=1/2$. 
Fig.~\ref{fig2} plots $\langle \sigma^z_{L/2}\sigma^z_{L/2+1}\rangle$  
and 
$\langle \sigma^z_{L/2}\sigma^z_{L/2+2}\rangle$ 
as a function of the time after the 
quench, and for initial thermal ensembles with $\beta=1/2,2$. In contrast to the Ising case, 
the stationary state at long times is expected to be described by a Gibbs ensemble. Since the energy is zero,  its temperature is infinite  (the Hamiltonian is traceless), and the expectation value of any traceless operator is zero. 
The data reported in Fig.~\ref{fig2} 
for $\beta=1/2$ are consistent with this picture:
after some initial oscillations, $\langle\sigma_{L/2}^z\sigma_{L/2+r}^z\rangle_c$ shows the tendency to
vanish. Instead, at lower temperatures, the approach to zero is much slower and, already for $\beta=2$, in the time window investigated the dynamics are practically the same as in the quench from the cat state \eqref{eq:cat} (dashed-dotted lines). 
The values of the prethermalization plateaux have been obtained as follows. We have considered small chains (up to $L=14$) and extrapolated the effective temperatures, $\beta_\Uparrow$ and $\beta_\Downarrow$, corresponding to the quenches from $\ket{\Uparrow}$ and $\ket{\Downarrow}$, respectively (imposing energy densities $E/L=\pm1/2$). Then, we numerically computed the arithmetic mean of the thermal expectation values (\emph{cf.} \eqref{eq:catGGE}).  These are shown as dashed horizontal lines in Fig.~\ref{fig2}. 
Clearly (assuming thermalization in the ANNNI model), the times reached are  too small to see the prethermalization plateaux. However, the picture is perfectly consistent with the mechanism presented: for sufficiently large $\beta$, the dynamics get stuck in a cat state, and cluster decomposition property is restored very slowly.

\paragraph*{Conclusions.}

We presented a new mechanism leading to prerelaxation and prethermalization  after quantum quenches in low dimensional quantum many-body systems.
We focused on one-dimensional models with magnetically ordered ground states. 
At small temperature and small length scales, the time-evolving state exhibits remnants of the zero-temperature 
long-range correlations, and violates cluster decomposition property. Precisely, 
in a typical time scale proportional to the finite-temperature correlation length,
the state is locally equivalent to a cat state. 
This happens both in integrable and in non-integrable 
post-quench dynamics. 
The prethermalization (prerelaxation) plateaux can be described by the arithmetic mean of Gibbs (generalized Gibbs) ensembles. 

Our work opens several interesting research directions. 
First, in the TFIC we identified a prerelaxation scaling limit where the dynamics occur in a time scale proportional to the thermal correlation length of the initial state - \emph{cf}. \eqref{pla-an}. The generality of this result has not been established, and, in the non-integrable case, could be, in fact, undermined by the slowness of the relaxation process of the emerging cat state. Further investigations in other models with a faster thermalization dynamics could help clarify this question.  
Second, in our scenario 
we considered the case of broken {\it discrete} symmetries. While zero-temperature 
long-range order with breaking of a continuous symmetry is forbidden in 1D, quasi-long-range order (QLRO) is possible. 
In the light of our findings, we wonder whether the presence of QLRO could give rise to similar prethermalization regimes. 
It would also be interesting to investigate the interplay between long-range 
order and prethermalization in higher dimensions, where even continuous symmetries can be broken. 
In this respect, prethermalization plateaux have been already observed in the relaxation dynamics from a spiral state in the 
3D Heisenberg model~\cite{babadi-2015}. 
Finally, it would be timely and of great interest  to consider this form of prethermalization in the presence of inhomogeneities, \emph{e.g.}, splitting the state in two, and preparing the two parts at different (small) temperatures (see, \emph{e.g.}, the very recent works \cite{CAD:hydro}), or introducing localized defects in the Hamiltonian~\cite{BF:16}.

\begin{acknowledgments}
We thank Luca Tagliacozzo for interesting discussions and for collaboration at an early stage of this project. We also thank Bruno Bertini and Pasquale Calabrese for a critical reading of the manuscript. 
This work was supported by  the European Union's Horizon 2020 research and innovation programme under the Marie Sklodowoska-Curie grant agreement No 702612 \emph{OEMBS} (VA), and  by LabEX ENS-ICFP:ANR-10-LABX-0010/ANR-10-IDEX-0001-02 PSL* (MF).
\end{acknowledgments} 



\begin{thebibliography}{99}

\bibitem{exp} M. Greiner,  O. Mandel, T.W. H\"ansch, and I. Bloch, 
Nature \href{\doi10.1038/nature00968}{\bf 419}, 51-54 (2002); T. Kinoshita, T. Wenger,  and D. S. Weiss, 
 Nature \href{\doi10.1038/nature04693}{\bf 440}, 900 (2006);
 S. Hofferberth, I. Lesanovsky, B. Fischer, T. Schumm, and J. Schmiedmayer, 
Nature \href{\doi10.1038/nature06149}{\bf 449}, 324-327 (2007); 
L. Hackermuller, U. Schneider, M. Moreno-Cardoner, T. Kitagawa \emph{et al}, 
Science \href{\doi10.1126/science.1184565}{\bf 327}, 1621 (2010);
S. Trotzky, Y.-A. Chen,  A. Flesch, I. P. McCulloch \emph{et al}, 
Nature Phys. \href{\doi10.1038/nphys2232}{\bf 8}, 325 (2012); M. Gring, M. Kuhnert, T. Langen, T. Kitagawa \emph{et al}, 
Science \href{\doi10.1126/science.1224953}{\bf 337}, 1318 (2012):
U. Schneider, L. Hackerm\"uller, J. P. Ronzheimer, S. Will, S. Braun \emph{et al}, 
Nature Phys. \href{\doi10.1038/nphys2205}{\bf 8}, 213 (2012); 
M. Cheneau, P. Barmettler, D. Poletti, M. Endres \emph{et al}, 
Nature \href{http://dx.doi.org/10.1038/nature10748}{\bf 481}, 484 (2012); 
T. Langen, R. Geiger, M. Kuhnert, B. Rauer, and J. Schmiedmayer,
Nature Physics \href{http://dx.doi.org/10.1038/nphys2739}{\bf 9}, 640 (2013);
F. Meinert, M.J. Mark, E. Kirilov, K. Lauber \emph{et al},  
Phys. Rev. Lett. \href{http://dx.doi.org/10.1103/PhysRevLett.111.053003}{\bf 111}, 053003 (2013); T. Fukuhara, A. Kantian, M. Endres, M. Cheneau \emph{et al}, 
Nature Physics \href{\doi10.1038/nphys2561}{\bf 9}, 235 (2013); T. Fukuhara, P. Schau{\ss}, M. Endres, S. Hild\emph{et al},  
Nature \href{\doi10.1038/nature12541}{\bf 502}, 76 (2013);
J.P. Ronzheimer, M. Schreiber, S. Braun, S.S. Hodgman \emph{et al}, 
Phys. Rev. Lett. \href{http://dx.doi.org/10.1103/PhysRevLett.110.205301}{\bf 110}, 205301 (2013);
P. Jurcevic, B. P. Lanyon, P. Hauke, C. Hempel \emph{et al}, 
Nature \href{http://dx.doi.org/10.1038/nature13461}{\bf 511}, 202 (2014).

\bibitem{EF:review} F.H.L. Essler and M. Fagotti, J. Stat. Mech. (2016) \href{\doi10.1088/1742-5468/2016/06/064002}{064002}.

\bibitem{GE15}
C. Gogolin and J. Eisert, Rep. Prog. Phys. \href{\doi10.1088/0034-4885/79/5/056001}{\bf 79}, 056001 (2016);
L. D'Alessio, Y. Kafri, A. Polkovnikov, and M. Rigol, 
Adv. Phys. \href{\doi10.1080/00018732.2016.1198134}{\bf 65}, 239 (2016).

\bibitem{quench}
M. Rigol, A. Muramatsu, and M. Olshanii,  
Phys. Rev. A \href{http://dx.doi.org/10.1103/PhysRevA.74.053616} {\bf 74}, 053616 (2006);
M. A. Cazalilla, 
Phys. Rev. Lett. \href{http://dx.doi.org/10.1103/PhysRevLett.97.156403} {\bf 97}, 156403 (2006);
P. Calabrese and  J. Cardy,  
J. Stat. Mech. (2007) \href{http://dx.doi.org/10.1088/1742-5468/2007/06/P06008}{P06008}; T. Barthel and U. Schollw\"ock, 
Phys. Rev. Lett. \href{http://dx.doi.org/10.1103/PhysRevLett.100.100601}{\bf 100}, 100601 (2008);
M. Cramer, C. M. Dawson, J. Eisert, and T. J. Osborne, 
Phys. Rev. Lett. \href{http://dx.doi.org/10.1103/PhysRevLett.100.030602}{\bf 100}, 030602 (2008);
A. Silva, Phys. Rev. Lett. \href{http://dx.doi.org/10.1103/PhysRevLett.101.120603}{\bf 101}, 120603 (2008);
S. Sotiriadis, P. Calabrese, and J. Cardy, EPL \href{\doi10.1209/0295-5075/87/20002}{\bf 87} (2009) 20002;
P. Calabrese, F. H. L. Essler and M. Fagotti,  
Phys. Rev. Lett. \href{http://dx.doi.org/10.1103/PhysRevLett.106.227203}{\bf 106}, 227203 (2011);
M. Fagotti and F.H.L. Essler, Phys.~Rev.~B \href{http://dx.doi.org/10.1103/PhysRevB.87.245107}{\bf 87}, 245107 (2013);
J.-S. Caux and R. M. Konik, 
Phys. Rev. Lett. \href{http://dx.doi.org/10.1103/PhysRevLett.109.175301}{\bf 109}, 175301 (2012);
F. H. L. Essler, S. Evangelisti, and M. Fagotti, 
Phys. Rev. Lett.  \href{http://dx.doi.org/10.1103/PhysRevLett.109.247206}{\bf 109}, 247206 (2012);
M. Collura, S. Sotiriadis, and P. Calabrese, 
Phys. Rev. Lett. \href{http://dx.doi.org/10.1103/PhysRevLett.110.245301}{\bf 110}, 245301 (2013);
J.-S.~Caux and F.H.L.~Essler, Phys. Rev. Lett. \href{http://dx.doi.org/10.1103/PhysRevLett.110.257203}{\bf 110}, 257203 (2013);
G. Mussardo, 
Phys. Rev. Lett. \href{http://dx.doi.org/10.1103/PhysRevLett.111.100401} {\bf 111}, 100401 (2013); 
B. Pozsgay, 
J. Stat. Mech. (2013) \href{http://dx.doi.org/10.1088/1742-5468/2013/07/P07003}{P07003};
M. Fagotti and F. H. L. Essler, 
J. Stat. Mech. (2013) \href{http://dx.doi.org/10.1088/1742-5468/2013/07/P07012}{P07012}; W. Liu and N. Andrei, Phys. Rev. Lett. \href{http://dx.doi.org/10.1103/PhysRevLett.112.257204}{\bf 112}, 257204 (2014);
J. De Nardis, B. Wouters, M. Brockmann, and J.-S. Caux, Phys. Rev. A \href{http://dx.doi.org/10.1103/PhysRevA.89.033601}{\bf 89}, 033601 (2014);
M. Fagotti, M. Collura,  F. H. L. Essler, and P. Calabrese,  
Phys. Rev. B \href{http://dx.doi.org/10.1103/PhysRevB.89.125101} {\bf 89}, 125101 (2014);
B. Wouters, J. De Nardis,  M. Brockmann, D. Fioretto \emph{et al},  
Phys. Rev. Lett.  \href{http://dx.doi.org/10.1103/PhysRevLett.113.117202}{\bf 113}, 117202 (2014);
B. Pozsgay, M. Mesty\'{a}n, M. A. Werner, M. Kormos \emph{et al},  
Phys. Rev. Lett. \href{http://dx.doi.org/10.1103/PhysRevLett.113.117203}{{\bf 113}}, 117203 (2014);
G. Goldstein and N. Andrei, Phys. Rev. A \href{\doi10.1103/PhysRevA.90.043625}{\bf 90}, 043625 (2014);
F. H. L. Essler, G. Mussardo, and M. Panfil,  
Phys. Rev. A \href{http://dx.doi.org/10.1103/PhysRevA.91.051602}{\bf 91}, 051602(R) (2015); 
L. Vidmar, D. Iyer, and M. Rigol, arXiv:\href{https://arxiv.org/abs/1512.05373}{1512.05373}.
S. Sotiriadis, Phys. Rev. A \href{\doi10.1103/PhysRevA.94.031605}{\bf 94}, 031605 (2016); L. Piroli, P. Calabrese, F.H.L. Essler, Phys. Rev. Lett. \href{\doi10.1103/PhysRevLett.116.070408}{\bf 116}, 070408 (2016); B. Bertini, L. Piroli, P. Calabrese, J. Stat. Mech. (2016) \href{\doi10.1088/1742-5468/2016/06/063102}{063102}; M. Kormos, M. Collura, G. Takács, P. Calabrese, Nature Phys. 21/09/2016 (\href{\doi10.1038/nphys3934}{doi:10.1038/nphys3934})
M.~Marcuzzi, J.~Marino, A. Gambassi, and A. Silva, Phys. Rev. B \href{https://doi.org/10.1103/PhysRevB.94.214304}{\bf 94}, 214304 (2016);
V. Alba and P. Calabrese, arXiv:\href{https://arxiv.org/abs/1608.00614}{1608.00614}; L. Piroli, E. Vernier, P. Calabrese, and M. Rigol, arXiv:\href{https://arxiv.org/abs/1611.08859}{1611.08859}.

\bibitem{SC:CD} S. Sotiriadis and P. Calabrese, J. Stat. Mech. (2014) \href{\doi10.1088/1742-5468/2014/07/P07024}{P07024}.

\bibitem{ETH}
J. M. Deutsch, Phys. Rev. A
\href{http://dx.doi.org/10.1103/PhysRevA.43.2046}{\bf 43}, 2046
(1991);   M. Srednicki, Phys. Rev. E
  \href{http://dx.doi.org/10.1103/PhysRevE.50.888}{\bf 50}, 888
  (1994);
  M.~Rigol, V.~Dunjko, and M.~Olshanii, Nature \href{http://dx.doi.org/10.1038/nature06838}{\bf 452}, 854 (2008). 
  
  
\bibitem{charges}
V.E. Korepin, A.G. Izergin, and N.M. Bogoliubov, {\em {Quantum Inverse
  Scattering Method, Correlation Functions and Algebraic Bethe Ansatz}}
  (Cambridge University Press, 1993); 
 T. Prosen, Nucl. Phys. B \href{
http://dx.doi.org/10.1016/j.nuclphysb.2014.07.024}{\bf 886} (2014) 1177; R.G. Pereira, V. Pasquier, J. Sirker, and I. Affleck, J. Stat. Mech. (2014) \href{\doi10.1088/1742-5468/2014/09/P09037}{P09037}; L. Piroli and E. Vernier, J. Stat. Mech. (2016) \href{http://dx.doi.org/10.1088/1742-5468/2016/05/053106}{053106}; E. Ilievski, M. Medenjak, and T. Prosen, Phys. Rev. Lett. \href{http://dx.doi.org/10.1103/PhysRevLett.115.120601}{\bf 115}, 120601 (2015); E. Ilievski, J. De Nardis, B. Wouters, J.-S. Caux \emph{et al},  
Phys. Rev. Lett. \href{http://dx.doi.org/10.1103/PhysRevLett.115.157201}{\bf 115}, 157201 (2015); E. Ilievski, M. Medenjak, T. Prosen, and L. Zadnik, J. Stat. Mech. (2016) \href{ttp://dx.doi.org/10.1088/1742-5468/2016/06/064008}{064008}; M. Fagotti, J. Phys. A: Math. Theor. \href{\doi10.1088/1751-8121/50/3/034005}{\bf 50} 034005 (2017); A. De Luca, M. Collura, and J. De Nardis, arXiv:\href{https://arxiv.org/abs/1612.07265}{1612.07265}. 
  
\bibitem{rigol-2007}
M.~Rigol, V.~Dunjko, V.~Yurovsky, and M.~Olshanii, Phys.\ Rev.\ Lett.\ 
\href{https://doi.org/10.1103/PhysRevLett.98.050405}{\bf 98}, 050405 (2007). 

\bibitem{preT} T. Langen, T. Gasenzer, and J. Schmiedmayer, J. Stat. Mech. (2016) \href{http://dx.doi.org/10.1088/1742-5468/2016/06/064009}{064009}.

\bibitem{prethermalization}
C. Kollath, A. M. L\"auchli, and E. Altman, 
Phys. Rev. Lett. \href{http://dx.doi.org/10.1103/PhysRevLett.98.180601}{\bf 98}, 180601 (2007);      
M. Moeckel and S. Kehrein, 
Phys. Rev. Lett. \href{http://dx.doi.org/10.1103/PhysRevLett.100.175702}{\bf 100}, 175702 (2008);
Ann. of Phys. \href{\doi10.1016/j.aop.2009.03.009}{\bf 324}, 2146 (2009); M Rigol, Phys. Rev. Lett. \href{\doi10.1103/PhysRevLett.103.100403}{\bf 103}, 100403 (2009);
 M. Kollar, F.A. Wolf, and M. Eckstein, 
Phys. Rev. B \href{http://dx.doi.org/10.1103/PhysRevB.84.054304}{\bf 84}, 054304 (2011);
M. Stark and M. Kollar,
arXiv:\href{http://arxiv.org/abs/1308.1610}{1308.1610} (2013);
M. Marcuzzi, J. Marino, A. Gambassi, and A. Silva, 
Phys. Rev. Lett. \href{http://dx.doi.org/10.1103/PhysRevLett.111.197203}{\bf 111}, 197203 (2013);
A. Mitra, 
Phys. Rev. B \href{\doi10.1103/PhysRevB.87.205109}{\bf 87}, 205109 (2013);
 A. Chiocchetta, M. Tavora, 
 A. Gambassi, and A. Mitra, 
Phys. Rev. B \href{http://dx.doi.org/10.1103/PhysRevB.91.220302}{\bf 91}, 220302(R) (2015); 
 B. Bertini, F.H.L. Essler, S. Groha, N.J. Robinson,  
Phys. Rev. Lett. \href{http://dx.doi.org/10.1103/PhysRevLett.115.180601}{\bf 115}, 180601 (2015); B. Bertini, F.H.L. Essler, S. Groha, N.J. Robinson, 
Phys. Rev. B \href{https://doi.org/10.1103/PhysRevB.94.245117}{\bf 94}, 245117 (2016).
G.P. Brandino, J.-S. Caux, and R.M. Konik, 
Phys. Rev. X \href{http://dx.doi.org/10.1103/PhysRevX.5.041043}{\bf 5}, 041043 (2015);
 N. Nessi and A. Iucci, 
arXiv:\href{http://arxiv.org/abs/1503.02507}{1503.02507}; A. Chiocchetta, A. Gambassi, S. Diehl, and J. Marino, arXiv:\href{https://arxiv.org/abs/1612.02419}{1612.02419}. 

\bibitem{babadi-2015}
M.~Babadi, E.~Demler, and M.~Knap, Phys.\ Rev.\ X \href{https://doi.org/10.1103/PhysRevX.5.041005}{\bf 5}, 
041005 (2015). 

\bibitem{essler-2014}
F.~H.~L.~Essler, S.~Kehrein, S.~R.~Manmana, and N.~J.~Robinson, Phys.\ Rev.\ B \href{https://doi.org/10.1103/PhysRevB.89.165104}{\bf 89}, 
165104 (2014).

\bibitem{preT:exp}
T. Kitagawa, A. Imambekov, J. Schmiedmayer, and E. Demler, New J. Phys. \href{http://dx.doi.org/10.1088/1367-2630/13/7/073018}{\bf 13}, 073018 (2011); 
M.~Gring, M.~Kuhnert, T.~Langen, T.~Kitagawa \emph{et al}, 
Science \href{\doi10.1126/science.1224953}{\bf 337}, 
6100 (2012); D. Adu Smith, M. Gring, T. Langen, M. Kuhnert \emph{et al}, 
New J. Phys. \href{http://dx.doi.org/10.1088/1367-2630/15/7/075011}{\bf 15} 075011 (2013).


\bibitem{prerelaxation}
M. Fagotti, 
J. Stat. Mech. (2014) \href{ttp://dx.doi.org/10.1088/1742-5468/2014/03/P03016}{P03016};
 B. Bertini and M. Fagotti, 
J. Stat. Mech. (2015) \href{http://dx.doi.org/10.1088/1742-5468/2015/07/P07012}{P07012};
 M. Fagotti and M. Collura, 
arXiv:\href{http://arxiv.org/abs/1507.02678}{1507.02678} (2015).

\bibitem{sachdev-book}
S.~Sachdev, {\it Quantum Phase Transitions}, Cambridge University Press (2011).


\bibitem{chakravarty-1989}
S.~Chakravarty, B.~I.~Halperin, and D.~R.~Nelson, Phys.\ Rev.\ B \href{https://doi.org/10.1103/PhysRevB.39.2344}{\bf 39}, 2344 
(1989).

\bibitem{Mermin} N.D. Mermin and H. Wagner, 
Phys. Rev. Lett. \href{https://doi.org/10.1103/PhysRevLett.17.1133}{\bf 17}, 1133 (1966).
\bibitem{Hohenberg} P.C. Hohenberg, 
Phys. Rev. \href{https://doi.org/10.1103/PhysRev.158.383}{\bf 158},
383 (1967).

\bibitem{bravyi06}
S. Bravyi, M. B. Hastings, and F. Verstraete, 
Phys. Rev. Lett.
\href{http://dx.doi.org/10.1103/PhysRevLett.97.050401}{\bf 97}, 050401 (2006).

\bibitem{LR72}
E. H. Lieb and D. W. Robinson, 
Commun. Math. Phys. \href{http://dx.doi.org/10.1007/BF01645779}{\bf 28}, 251
(1972).

\bibitem{f:5}
The relaxation time can be defined as the time after which a distance between the reduced density matrix and its stationary value is smaller than a given $\epsilon$. 

\bibitem{localT} M. Kliesch, C. Gogolin, M.J. Kastoryano, A. Riera, and J. Eisert, Phys. Rev. X \href{\doi10.1103/PhysRevX.4.031019}{\bf 4}, 031019 (2014); S. Hernández-Santana, A. Riera, K.V. Hovhannisyan, M. Perarnau-Llobet \emph{et al}
, New J. Phys. \href{http://dx.doi.org/10.1088/1367-2630/17/8/085007}{\bf 17} 085007 (2015).

\bibitem{weinberg-book}
S.~Weinberg, {\it The quantum theory of fields}, Cambridge University Press (1995).


\bibitem{tDMRG} S. R. White and A. E. Feiguin, Phys. Rev. Lett. \href{https://doi.org/10.1103/PhysRevLett.93.076401}{\bf 93}, 076401
(2004); A. J. Daley, C. Kollath, U. Schollw\"ock, and G. Vidal,
J. Stat. Mech. (2004) \href{http://dx.doi.org/10.1088/1742-5468/2004/04/P04005}{P04005}.

\bibitem{f:2} In the numerical simulations, we imposed open boundary conditions, and focused on observables in the bulk. 

\bibitem{f:4} If the model is integrable we use the term prerelaxation instead of prethermalization. 

\bibitem{f:1}
If $\gamma r$ is not small, the r.h.s. of \eqref{pla-an} has the corrective term
$$
\iint\limits_{[-\pi,\pi]^2}\frac{\mathrm d k\mathrm d p}{(2\pi)^2}\cos^2\theta_p\cos^2\theta_k\sum_{\sigma=-1,1}\frac{e^{-\gamma |(\varepsilon'_p-\varepsilon'_k) t+\sigma r|}}{2}\, .
$$


\bibitem{CAD:hydro}
A. Biella, A. De Luca, J. Viti, D. Rossini \emph{et al}, 
Phys. Rev. B \href{https://doi.org/10.1103/PhysRevB.93.205121}{\bf 93}, 205121 (2016);
O. A. Castro-Alvaredo, B. Doyon, and T. Yoshimura, Phys. Rev. X \href{https://doi.org/10.1103/PhysRevX.6.041065}{\bf 6}, 041065 (2016);
B. Bertini, M. Collura, J. De Nardis, and M. Fagotti,  
Phys. Rev. Lett. \href{\doi10.1103/PhysRevLett.117.207201}{\bf 117}, 207201 (2016). 

\bibitem{BF:16} B. Bertini and M. Fagotti, Phys. Rev. Lett. \href{\doi10.1103/PhysRevLett.117.130402}{\bf 117}, 130402 (2016) 

\end{thebibliography}
\end{document}